# A bright impulsive solar burst detected at 30 THz

P. Kaufmann[1,2,*], S. M. White[3], S. L. Freeland[4], R. Marcon[5,6], L.O.T. Fernandes[1], A.S. Kudaka[1], R.V. de Souza[1], J.L. Aballay[7], G. Fernandez[7], R. Godoy[7], A.Marun[7], A.Valio[1], J.-P. Raulin[1], C.G. Giménez de Castro[1]

[1]Center of Radio Astronomy and Astrophysics, Engineering School, Mackenzie Presbyterian University, São Paulo, SP, Brazil.

[2]Center of Semiconductor Components, State University of Campinas, Campinas, SP, Brazil.

[3]Air Force Research Laboratories, Space Vehicles Directorate, Albuquerque, NM 87117, USA.

[4]Lockheed Martin Solar and Astrophysics Laboratory, Palo Alto, California 94304, USA.

[5]"Gleb Wataghin" Physics Institute, State University of Campinas, Campinas, SP, Brazil.

[6]"Bernard Lyot" Solar Observatory, Campinas, SP. Brazil.

[7]El Leoncito Astronomical Complex, CONICET, San Juan, Argentina.

**Abstract**

Ground- and space-based observations of solar flares from radio wavelengths to gamma-rays have produced considerable insights but raised several unsolved controversies. The last unexplored wavelength frontier for solar flares is in the range of submillimeter and infrared wavelengths. Here we report the detection of an intense impulsive burst at 30 THz using a new imaging system. The 30 THz emission exhibited remarkable time coincidence with peaks observed at microwave, mm/submm, visible, EUV and hard X-ray wavelengths. The emission location coincides with a very weak white-light feature, and is consistent with heating below the temperature minimum in the atmosphere. However, there are problems in attributing the heating to accelerated electrons. The peak 30 THz flux is several times larger than the usual microwave peak near 9 GHz, attributed to non-thermal electrons in the corona. The 30 THz emission could be consistent with an optically thick spectrum increasing from low to high frequencies. It might be part of the same spectral component found at sub-THz frequencies whose nature remains mysterious. Further observations at these wavelengths will provide a new window for flare studies.

**1. Introduction**

One of the few relatively unexplored wavelength ranges for solar flare studies is the mid-infrared (i.e., THz frequencies). Recent discoveries in the emission of solar flares at sub-THz frequencies have motivated more interest in this range: a number of solar bursts have been observed to have an unexpected component in the sub-THz range that is related to but distinct from the well-known non-thermal microwave emission from accelerated electrons that has its maximum typically at a few tens of GHz (Kaufmann *et al.* 2004; 2006; Silva *et al.* 2007).

The emission component increasing through sub-THz frequencies poses fundamental problems for interpretation, and in turn raises the question of the behavior of solar flares at poorly-observed THz frequencies (Fleishman & Kontar 2010). Most observations in the upper THz frequency range observable from the ground in the mid-infrared band (8-15 µm, or 30 THz) have focused on quiet and quiescent regions (Turon & Léna 1970; Lindsey & Heasley 1981; Gezari, Livingston & Varosi 1999; Marcon et al. 2008, Cassiano et al. 2010) and magnetic field measurements in active regions using Zeeman spectroscopy (Jennings et al. 2002; Sonnabend et al. 2006), with little attention paid to the continuum emission.

Here we report the observation of an intense solar flare continuum at 30 THz (10 µm wavelength) obtained with an optical setup and camera designed for solar observations at El Leoncito observatory in the Argentinean Andes (Marcon et al. 2008; Kaufmann et al. 2008). We supplement this with high cadence EUV and white light images from the Solar Dynamics Observatory (SDO; NASA 2009), radio observations from the USAF RSTN network (Guidice 1979), spectrograph observations at 5-1200 MHz by the Green Bank Solar Burst Spectrometer (GBSRBS; Bastian et al. 2005), the millimeter-wave radio polarimeters (45 and 90 GHz; Valio et al., 2013) and the Solar Submillimeter Telescope (SST, 200 and 400 GHz; Kaufmann et al. 2008) located at El Leoncito, GOES soft X-rays (NOAA 2012), FERMI hard X-ray (HXR) measurements (Share & Murphy 2007) and RHESSI hard X-ray images that are available just after the impulsive phase (Lin et al. 2002). Observations were complemented by high cadence Hα solar flare imaging obtained at the SST site.

## 2. 30 THz observations

The 30 THz setup has been operating at El Leoncito, utilizing a coelostat with a 30 cm flat mirror, projecting the solar radiation into the instrumentation room, where it is reflected by another flat mirror into a 15 cm diameter Newtonian telescope. The arrangement is combined with a germanium lens to form an image of nearly ¾ of the solar disc at the focal plane 320 x 340 pixel microbolometer array of a FLIR A20 8-15 µm (30 THz) camera (Marcon et al. 2008; Kaufmann et al. 2008). To ensure that no near-IR and visible radiation contaminates the images, transmission tests were performed using specially designed > 3µm transmission filters (Tydex 2011). Frames were obtained at 5/s, and integrated over one second. Fig.1 shows a series of 30 THz frames exhibiting the intense burst brightening to the east of the leading spot in AR11429, peaking at about 17:23 UT.

Photometry is derived from the 30 THz images using the IRIS processing software (Buil 1999) and is illustrated in Fig. 2. The 30 THz time profile shown in Figure 3 represents the intensity in the inner 24" circle A after subtracting a photospheric level measured from the outer ring B, 30-62". The observed 30 THz excess temperature has been calibrated from measured units to brightness temperature by assuming a 4700 K difference between the effective temperatures of the photosphere and the sky, assuming the disk temperature at 30 THz to be about 5000 K (Turon & Léna 1979) and the sky to be at about the ambient temperature of 300 K (due to the mirrors and blocking devices interposed). This scale proved to be approximately correct, since it provides a sunspot temperature nearly 450 K less than the photosphere, which is consistent with measurements by other authors (Turon & Léna 1979; Gezari, Livingston &Varosi 1999; Marcon et al. 2008). We assume that there is an uncertainty of order 10% in the solar disk temperature at 30 THz, depending on the atmospheric model used. The 30 THz flux density may be assigned assuming that the burst size is smaller than the point-spread function defined by the 15 cm aperture diffraction-limited angle, i.e., about 15 arcseconds, and that the Rayleigh-Jeans approximation to Planck's law is still valid at 30 THz (Phillips 1987). The aperture efficiency has been estimated taking into account the fact that the instrument assembly has five reflections on aluminized mirrors, each transmitting about 0.8, plus the 0.46 transmission by the 4 mm thick Ge lens (Tydex 2012), and 0.85 transmissions due to the physical blockage at the telescope, resulting in a final aperture efficiency of 0.13, and effective aperture of 0.0023 m$^2$. The flux density is given by the well known expression (Kraus 1986) $\Delta S = 2 k \Delta T/A_e$, where k is the Boltzmann constant, $A_e$ the effective area corresponding to 0.0023 m$^2$, and $\Delta T = 100$ K is the observed 30 THz excess temperature at the peak. This correspond to a flux of 12000 SFU (with an uncertainty of order 25%, i.e. ± 3000 SFU; 1 SFU =$10^{-22}$ Wm$^{-2}$Hz$^{-1}$).

### 3. The 30 THz burst position

The event was a GOES class M8 flare occurring on March 13, 2012 in NOAA active region 1429 at heliographic coordinates N18W50 (on March 12, 24:00 UT, NOAA 2012). The 30 THz photometry is shown in the top panel of Fig. 3, applying a running mean of 10 data points to frames taken with a cadence of 0.61 s. Superimposed is the white-light time profile derived from SDO/HMI images every 45 seconds. The bottom panel, left side, shows the 30 THz image at a time close to the impulsive peak. The 30 THz spatial position accuracy is of about 10 arcseconds. Note that there is pronounced limb darkening at 30 THz and the apparent limb in these images is well inside the photospheric limb. In the lower right side of Fig. 3 we show (upper panel) the white light image of the region, (middle panel) a white-light difference image produced by subtracting a preflare image from images near 17:24 UT, and (lower panel) the SDO AIA 1600 Å image at 17:31 UT (after the impulsive phase to minimize saturation). The 30 THz burst position coincides with the eastern white-light and 1600 Å brightening.

SDO/AIA images show a great deal of activity in and around the flare site over an hour leading up to the M8 flare at 17:20 UT. Fig. 4 shows selected SDO EUV 211 Å images. A bright arch with a height of order 20000 km is visible over the flare site from about 16:45 onwards due to an earlier energy release. This arch is remarkably stable during the flare. In the lower atmosphere two bright ribbons are seen in most of the AIA bands (see Figs. 4 and 5), one extending south from the west side of the leading spot, the other extending north and east from a location 30 arcseconds east of the leading spot. The 30 THz source location corresponds to the brightest emission feature seen in the eastern ribbon, close to its western terminus and to the southern footpoint of the pre-existing bright arch. During the impulsive phase hot coronal loops are seen to connect the western end of the eastern ribbon and the eastern end of the western ribbon, and this develops into a post-flare loop system after 17:33 UT (Fig. 5).

This event was associated with a coronal mass ejection whose speed was estimated to be of order 1900 km/s (note that the leading edge of the CME was only present in a single LASCO C2 frame and this caused problems for CME catalogs that rely on C2 images), as well as a pronounced EUV wave and a large energetic particle event. At around 17:00 UT, tall loops above the active region are seen to start to expand slowly (speed of order 50 km/s) in the coronal EUV images. The expansion is seen to accelerate significantly at around 17:17 UT to about 140 km/s, followed by a dramatic eruption at around 17:22 UT, at which point the outer loops leave AIA's field of view (and their speed therefore cannot be estimated). We cannot see any clear evidence for an active-region filament eruption in conjunction with this event at the available EUV, UV or Hα wavelengths. The pre-existing bright arch remains in place, while the post flare loop system has its eastern footpoints concentrated around the 30 THz source region but with the western footpoints spread out along a north-south ribbon to the west. Thus although the eastern ribbon appears to brighten in the AIA 1600 Å images of the chromosphere, there is no sign of coronal loops connecting this ribbon to the ribbons to the east until the post-flare phase (Fig. 5). Furthermore, the fact that the pre-existing bright coronal arch shows no disruption from the flare also suggests that the coronal response to the flare is confined to regions to the west of the arch. An interesting feature is that during the postflare phase, the southern leg of the bright preflare loop lies in front of lower hot flare loops and clear absorption features are seen in the EUV images (e.g., 211 Å in Fig. 5).

### 4. Time profiles and spectra

Time profiles for several diagnostics during the main part of the flare are shown in Fig. 6. This event was detected in high-energy hard X-rays by the FERMI satellite. The 30 THz time profile is consistent with the impulsive profiles obtained at microwaves by RSTN, with the 50-100 and 300-540 keV ranges from FERMI, the 45 and 90 GHz solar polarimeters and at 212 GHz. At 405 GHz the flux is below the detection limit. No impulsive burst was observed in Hα: the Hα photometry shows a slow intensity growth rate reduction, with a peak in the derivative, at the same time as the first peak of the impulsive burst.

The dynamic spectrum of the flare from 5-1200 MHz obtained by GBSRBS is shown in Fig. 7. Bright decimetric continuum is seen in the impulsive phase, starting at around 17:18 UT, and Type III-like features are seen at metric wavelengths. There is no unambiguous trace of a Type II burst in this spectrum.

In the microwave spectrum, six nearly regular pulses with a period of 1.2 minutes are apparent at 5.0 and 8.8 GHz before the peak at about 17:23 UT after a lower envelope is subtracted from the time profile (Fig. 8). No more pulses are seen afterwards. The last peak coincides in time with the 30 THz peak and the first peaks at 90, 45 and 212 GHz. It also has a strikingly good time coincidence with the 610 MHz narrow band spike, as is better demonstrated in Fig. 8, bottom panel.

The electromagnetic spectrum for the whole range of frequencies can be derived from the flux data obtained above, shown in Fig. 9. This is the spectrum at the main peak of the impulsive phase. The 30 THz detection is clearly distinct from the typical nonthermal radio burst with a peak at about 9 GHz, which is usually attributed to gyrosynchrotron emission from mildly relativistic electrons (Dulk 1985). A spectral index close to 2 connects the upper limit at 405 GHz (smaller than 10 SFU) and the 30 THz measurement (about 12000 SFU), consistent with optically thick emission that could be either thermal or non-thermal.

### 5. Emission mechanisms at 30 THz

Suggested mechanisms for IR flare continuum include synchrotron radiation from energetic electrons, thermal emission from the lower atmosphere, thermal free-free emission from dense coronal plasma, and thermal emission from the chromospheric region that produces flare Hα emission (Ohki & Hudson 1975). Atmospheric models place the optical depth $\tau=1$ layer for 30 THz emission at a height above the photosphere of order 150 km where the temperature is around 5000 K, between the photosphere and the temperature minimum (e.g.,Turon & Lena 1979; Fontenla, Balasubramanian & Harder 2007). The pronounced limb darkening at 30 THz evident in Figs. 1, 2 and 3 confirms that $\tau=1$ for 30 THz lies in a region of the atmosphere with a large negative vertical temperature gradient, so that the oblique path of a ray at the limb pushes the effective $\tau=1$ layer at 30 THz higher in the atmosphere where the temperature is lower.

The display range of the 30 THz image used in Fig. 3, which is intended to emphasize features on the disk, exaggerates the actual extent of limb darkening. Earlier studies, e.g., Lindsey & Heasley 1981, argued for little limb darkening at mid-IR wavelengths. However, Léna (1970) found limb darkening at 30 THz for heliocentric longitudes larger than 80$^o$. These facts together with our images suggest that the darkening is confined to the region close to the limb. The location of the 30 THz τ=1 layer provides a constraint that we can use to investigate the 30 THz emission mechanism: the τ=1 layer is so close to the optical photosphere that if the 30 THz emission arises there rather than in the corona, it suggests that white light emission might also be produced. Recently the height of the white-light continuum in a flare has been measured and found to be 195 ± 70 km above the photosphere (i.e., the 5000 Å τ=1 level), which is similar to the expected height of the 30 THz layer (Martinez Oliveros *et al.* 2012).

As shown earlier (Fig. 3), preflare-subtracted SDO/HMI white-light images during the flare do indeed show faint white-light emission visible at the peak of the impulsive phase (from 17:23-17:26 UT). Fig. 10 shows images of the lower atmosphere with the location of the white-light emission, compared to the SDO magnetogram and 1700 Å. The white-light emission is seen in two compact regions, one at the west end of the eastern ribbon, the other at the eastern end of the western ribbon. These are also the regions of brightest emission in the AIA UV images of the chromosphere (1600, 1700 Å) and in the coronal EUV images as well. The location of the 30 THz source (Fig. 3) is consistent with the eastern white-light source, and when the HMI images are convolved to the resolution of the 30 THz images, the eastern white-light source appears much brighter than the western source. However, in terms of increase in brightness, the western source has a larger relative increase (of order 20% over the preflare level) due to the fact that it lies over the penumbra, while the eastern source lies outside the sunspot and shows only an increase of order 10% over the preflare level. (Note that HMI's bandpass at around 6170 Å is relatively red: since white-light flares are generally blue in apparent color, e.g., Neidig 1989, larger contrast would be expected at shorter wavelengths.) Inspection of movies of the white-light emission shows small motions (several arcseconds) of the two sources over several minutes, in the directions indicated in Figs. 3 and 10. These directions are consistent with motion along the flare ribbons; such motions are also evident in the brightest features in the AIA UV chromospheric images.

The presence of white light emission at the location of the 30 THz source strongly suggests that in this event the 30 THz emission originates low in the solar atmosphere and not in the corona.

### 6. Discussion

If the 30 THz flare emission indeed arises close to the τ=1 layer, then its interpretation suffers the same difficulties as white-light emission: when observed, the white-light/UV continuum provides the bulk of the radiated energy in a flare and therefore requires that the bulk of the flare energy be deposited in the radiating layers, but the 10-50 keV electrons that are believed to carry the bulk of the flare energy lose that energy through collisions in the chromosphere and cannot reach the white-light photosphere (e.g.,Neidig 1989; Machado, Emslie & Avrett 1989) . Only extremely energetic particles (e.g., electrons of order 1 MeV) can penetrate to the depths required to produce white-light continuum. The FERMI data showing HXR above 300 keV indicate that MeV electrons were present in this event, but a fit to the FERMI data gives a photon spectral index of around 3.8 above 100 keV in the impulsive phase, corresponding to a thick-target electron energy spectral index of order 5.3, i.e., not particularly hard, and certainly not hard enough for MeV-electrons to dominate the flare energy budget.

An alternative mechanism for white-light emission is that the energy deposited in the chromosphere is re-radiated downwards at wavelengths corresponding to Balmer transitions where the lower chromosphere is transparent and thus "back-warms" the deeper layers where the optical continuum can be produced (Machado, Emslie &Avrett 1989). However, direct heating of the layers that produce the Balmer continuum by electron energy deposition is not favored in this mechanism, but rather irradiation by XUV photons produced higher in the atmosphere: hence this mechanism involves several stages of energy conversion, each of which requires high efficiency if it is to explain the large radiative energy output in the optical continuum.

A possible reconciliation of these facts is that the 30 THz emission is optically thick thermal emission and its flux is therefore non-linearly related to the rate of energy deposition. If weak acceleration of MeV-energy electrons takes place in the corona and results in sufficient heating of the lower atmosphere before FERMI can detect HXR, it could explain the slow early rise of 30 THz emission, but if so, given the orders of magnitude increase in 300-540 keV HXR after 17:21 UT compared to the factor of 3 increase in 30 THz flux, we can expect 30 THz observations to provide a very sensitive means for detecting the presence of such energetic electrons. We note that the 30 THz emission has its own characteristic properties and is not simply a proxy for white-light emission: in addition to the fact that a slow rise in 30 THz emission is seen several minutes before FERMI detects the onset of hard X-rays at energies that MeV electrons produce, the 30 THz emission also continues to be observable much longer than either the white-light or the 300-540 keV HXR emission. The impulsive 30 THz rise, however, coincides with hard X-rays, suggesting that there might be two 30 THz emission components superimposed.

The position of the 212 GHz burst centroid of emission has been verified using the method based on the comparison of relative excess antenna temperatures from the SST multiple beams technique (Georges *et al.* 1989; Giménez de Castro *et al.* 1999). It fits fairly well the 30 THz burst site position, within the expected ± 30 arcseconds uncertainty. The RHESSI X-ray imager (Lin *et al.* 2002) missed the impulsive phase of the burst. However the X-ray image obtained later (17:35 UT) show that the brightest source in harder X-rays (> 25 keV) is at a location that coincides with the 30 THz and eastern white-light position (see Fig. 11). This confirms that nonthermal particles are conveying energy at least to the chromosphere directly above the 30 THz site, but does not distinguish between the deep-precipitation and backwarming models for the optical/IR continuum emission.

No sub-THz component with a rising spectrum was detected in this event, but the presence of the 30 THz emission raises the question of whether the sub-THz components can be attributed to the same source, which we infer to be heating of the lower atmosphere. This possibility has been considered in the context of other events: here we infer an input temperature increase of order 100 K for the 30 THz emission, corresponding to a flux of 12000 SFU. For a source size of $\leq 10$ arcseconds, this implies a local brightness temperature increase of order 2000 K. But for the large fluxes ($> 10^3$ SFU) observed in other sub-THz events with source sizes of order 10 arcseconds, optically thick temperatures over $10^7$ K are required (e.g., Trottet *et al.* 2011), which is not consistent with a location in the lower atmosphere. Further, in the case of the bright white-light emission detected in the large 2006 Dec 06 flare, an extension to 405 GHz results in a flux of just a few SFU (Krucker *et al.* 2013). Based on these arguments it seems unlikely that the 30 THz emission in this event is relevant to understanding the bright sub-THz component reported in other events.

### 7. Concluding remark

The 30 THz bright solar burst detection clearly opens a promising new window for the study of flares. In this event the 30 THz emission was much cleaner than the white-light detection, allowing us to study the role and energetics of continuum from below the temperature minimum (which is known to dominate the radiative energy output of flares) as well as potentially providing a sensitive means of detecting the presence of MeV electrons, and in flares with a strong sub-THz component we may also expect additional 30 THz sources. The complete diagnostics, however, require other data at frequencies in the range of sub-THz to 30 THz.

**Acknowledgments**: This research was partially supported by Brazilian agencies FAPESP,CNPq, Mackpesquisa, Argentina CONICET, and US AFOSR and NASA. SW thanks Rachel Hock for valuable comments.

**Captions to the Figures**

**Fig. 1**. Sequence of 30 THz frames showing the burst brightening up east of the sunspot.

**Fig. 2**. The photometry technique subtracts the inner circle intensity on the burst source from the outer ring on the photosphere.

**Fig. 3**. The top panel shows the 30 THz and white-light time profiles of the impulsive solar burst observed March 13, 2012, obtained with 0.61 and 45 second cadence, respectively. The 30 THz ordinates are in degrees K in a scale calibrated using the photosphere – outside sky temperature difference (described in the text). The bottom images show at left the 30 THz brightening at the beginning of the impulsive rise, at 17:20:48 UT with a cross indicating the center of emission. The display range in this image is scaled to emphasize disk features and exaggerates the amount of limb darkening. At right the white light image of the flare region, top (from SDO/HMI), the white-light difference image produced by subtracting a preflare image from images summed around 17:24 UT (middle), and a 1600 Å image (bottom, from SDO/AIA at 17:31 UT) showing the flare ribbons. The solid contours represent the two white-light features seen in the middle right panel. Arrows indicate the 30 THz burst position that coincides with the eastern features at white-light and 1600 Å.

**Fig. 4.** Selected SDO 211 Å images of the flare. The first panel shows a tall activated magnetic arch existing before the burst.The first intense brightening is seen at 17:21:26 as the impulsive phase starts, and the brightening of the western feature is seen at 17:23:12. The small cross indicates the 30 THz burst position. The image in the decay phase at 17:31:00 UT shows that the big preflare arch remained steady in front of the flaring activity.

**Fig. 5.** SDO/AIA images of the flaring region before (upper panels) and after (lower panels) the impulsive phase at 1600 Å (upper chromosphere) and 211 Å (2 MK corona). The two flare ribbons are clearly visible in the later 1600 Å image, while the bright preflare arch is apparent in both 211 Å images. The two contours are the locations of the faint white-light emission and the brightest emission at all wavelengths: the 30 THz source location is consistent with the eastern contour. Note the development of hot flare loops connecting the ends of the two ribbons, and the absorption features in the later 211 Å image where the bright arch passes above lower flare loops.

**Fig. 6.** The comparison of time profiles for the March 13, 2012 solar burst, at 30 THz, with FERMI hard X-rays and RSTN 8.8 GHz microwaves in the top panel, and zoomed out at the bottom for the RSTN 8.8 GHz compared to the 30 THz, millimeter and submillimeter wave observations during the impulsive phase.

**Fig. 7.** The radio dynamic spectrum below 1200 MHz from the Green Bank Solar Radio Burst Spectrometer

**Fig. 8.** The 8.8 GHz burst with the lower envelope removed, and the intense narrowband spike at 610 MHz

**Fig. 9.** Radio to mid-IR spectrum for the March 13, 2012 solar burst.

**Fig. 10.** Images of the flare region in white light (upper left, from SDO/HMI), line-of-sight magnetogram (lower left, from SDO/HMI), and at 1700 Å (lower right, from SDO/AIA at 17:31 UT, after the impulsive phase to minimize saturation). The upper right panel shows a white-light difference image produced by subtracting a preflare image from images summed around 17:24 UT. The solid contours represent the extent of two white-light features seen in the data; arrows represent the directions of motion of the two white-light features seen over several minutes.

**Fig. 11.** RHESSI X-ray image obtained later (17:35 UT) show that the harder X-rays (25-50 keV, full line contours) come from a location that coincides with the 30 THz and eastern white-light position. Emission at lower energies (12-25 keV, dashed line contours) are not correlated in space.

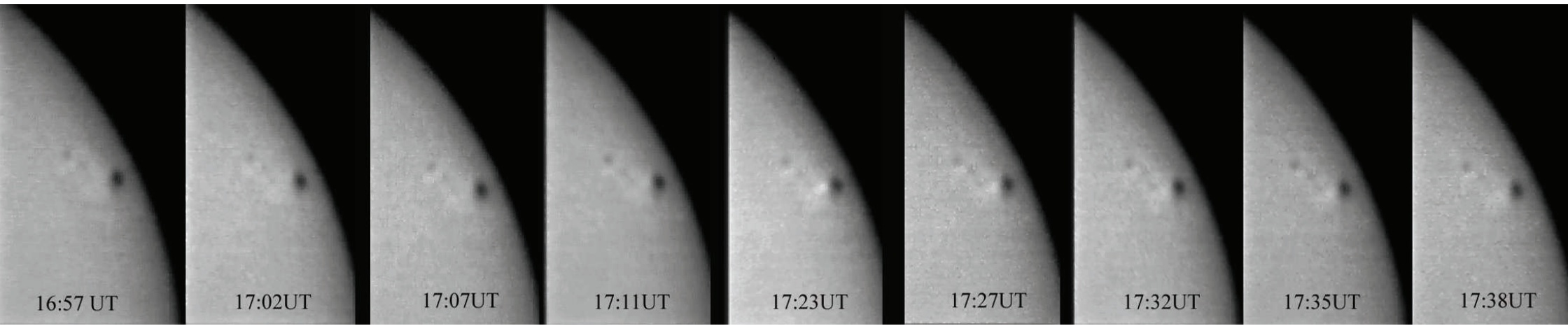

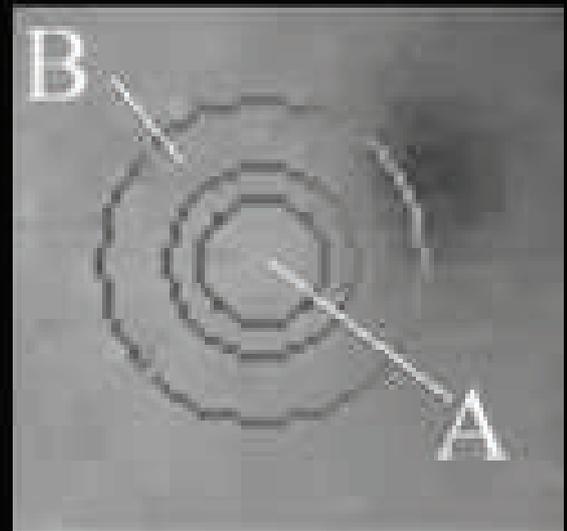

rA=8px
rintB=12px
rextB=20px

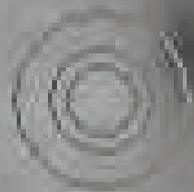

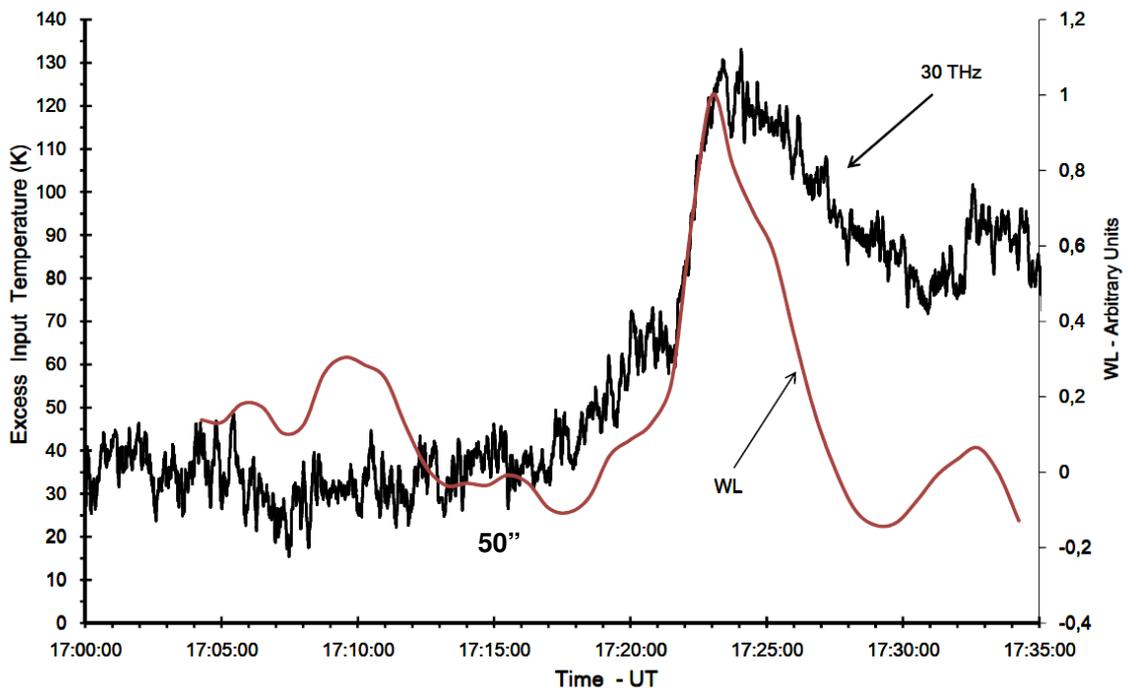
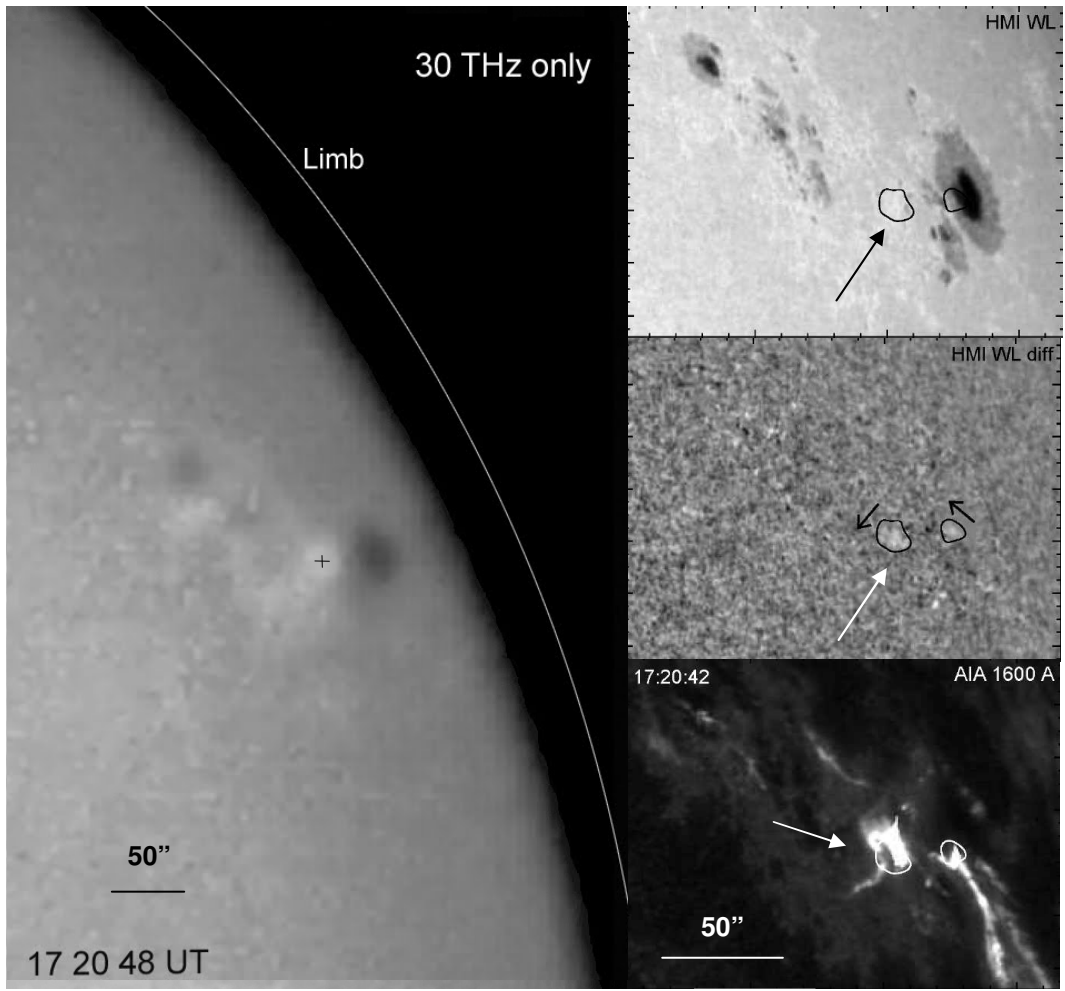

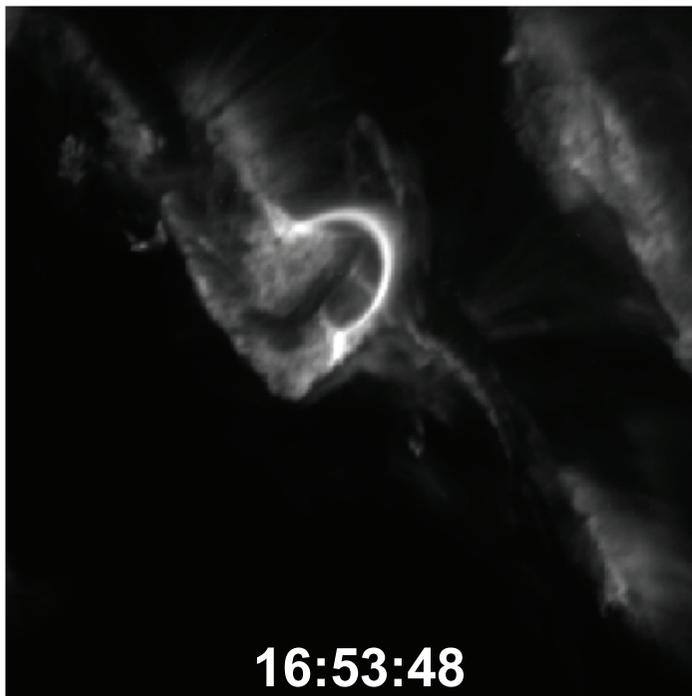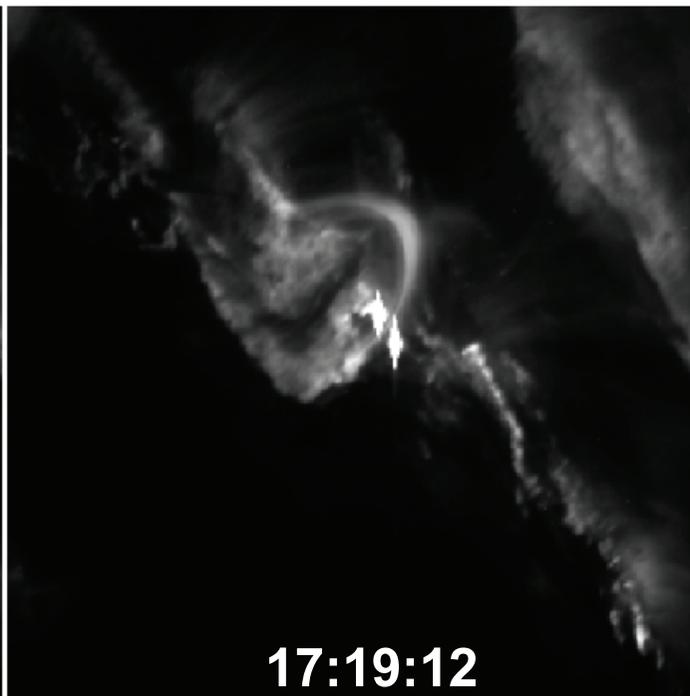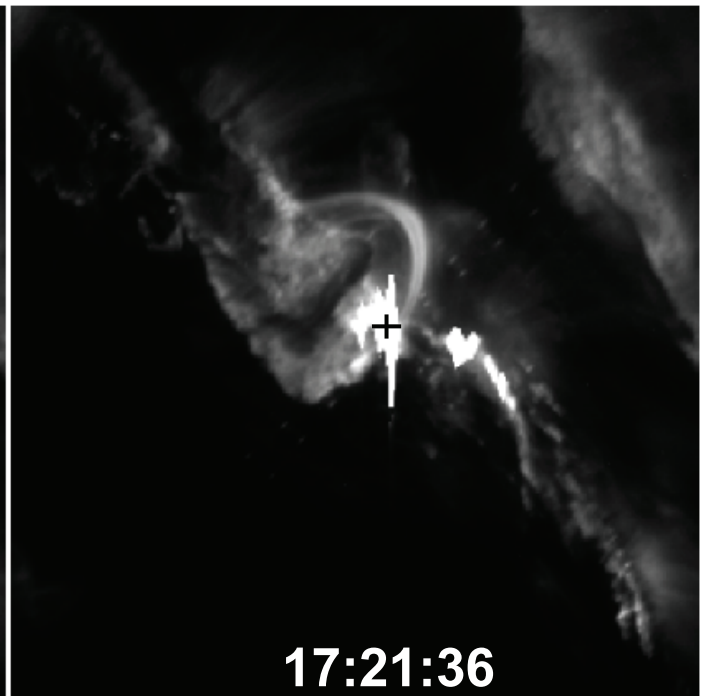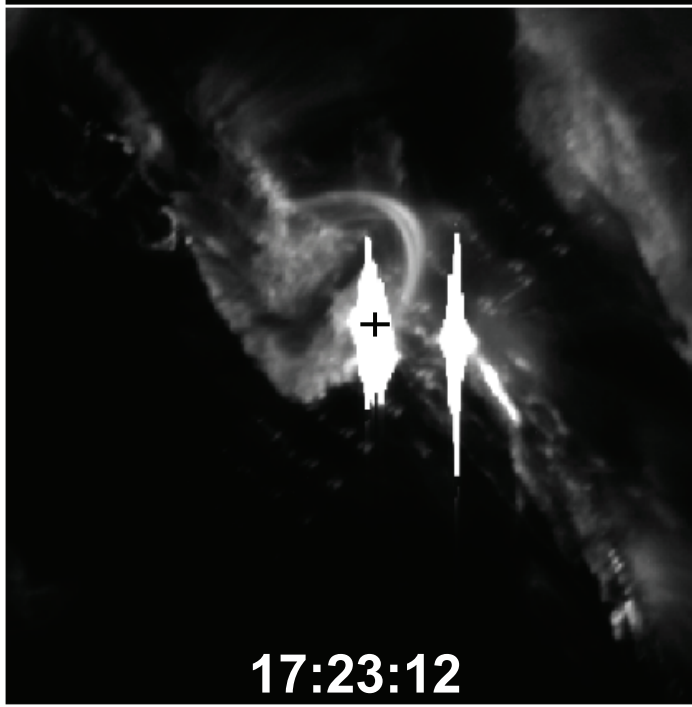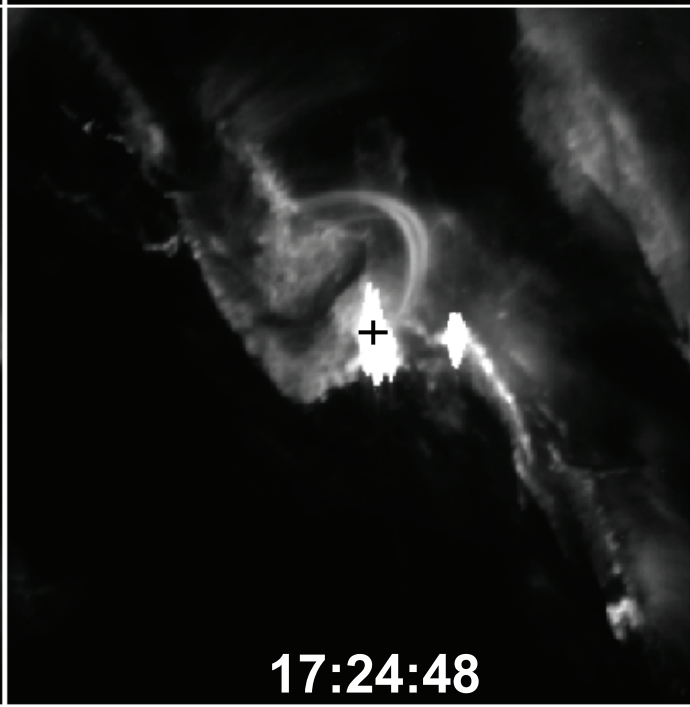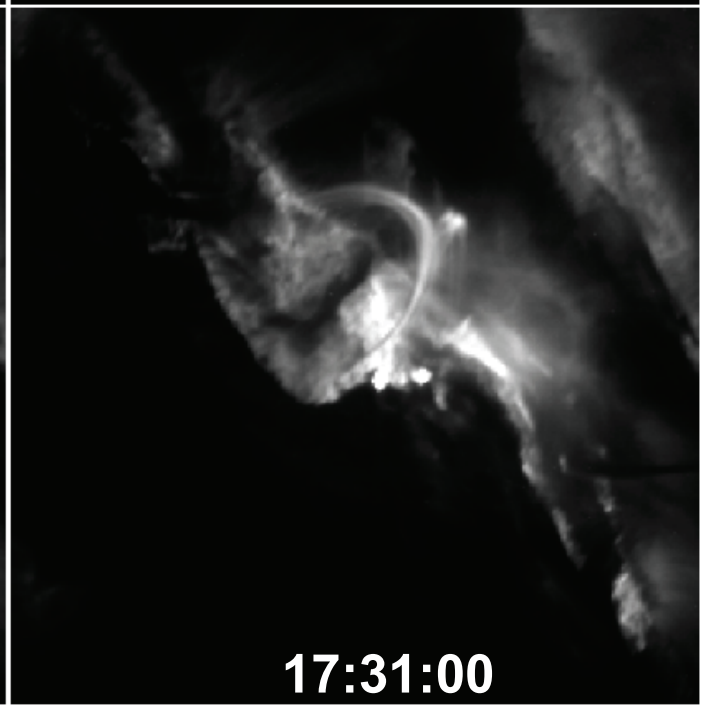

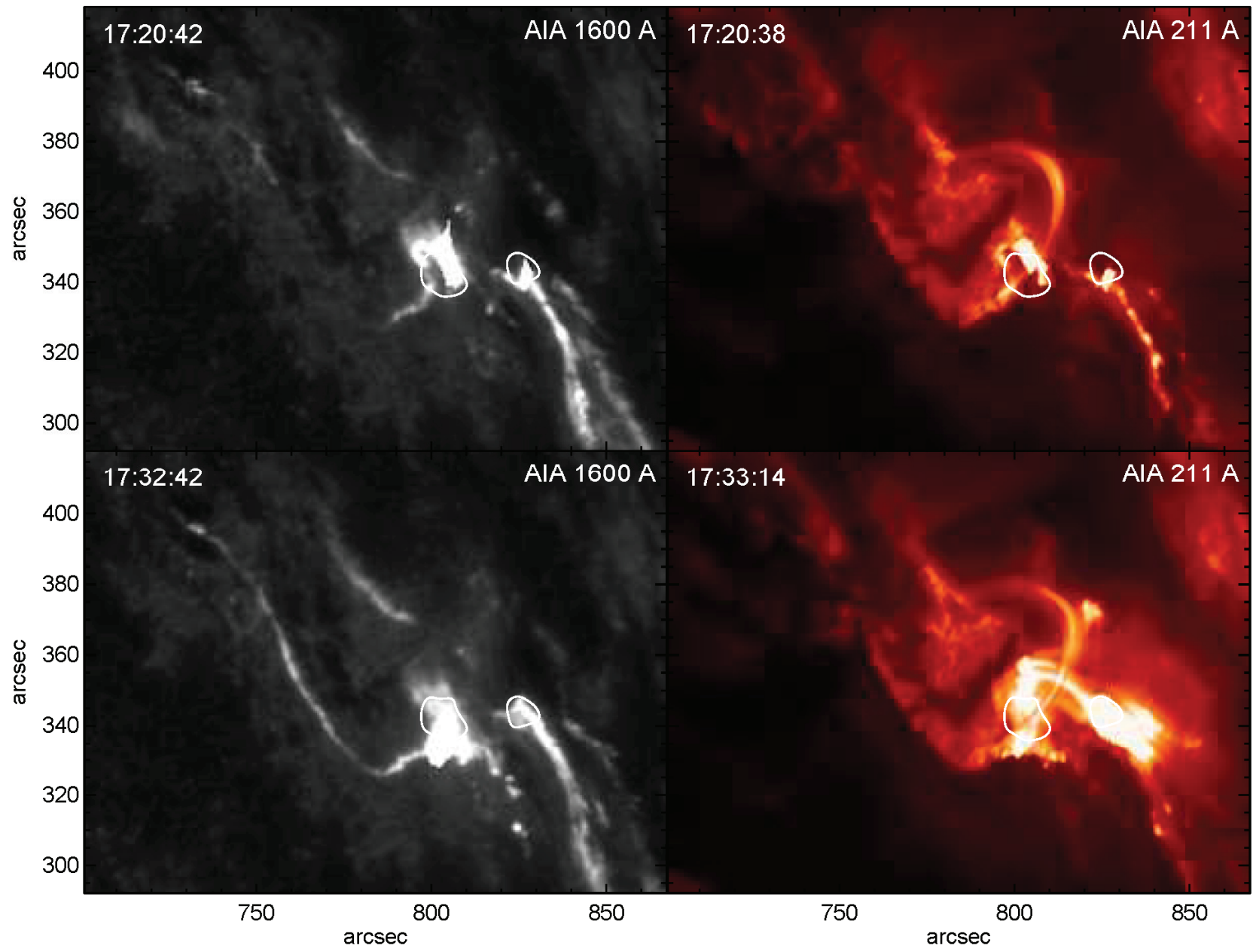

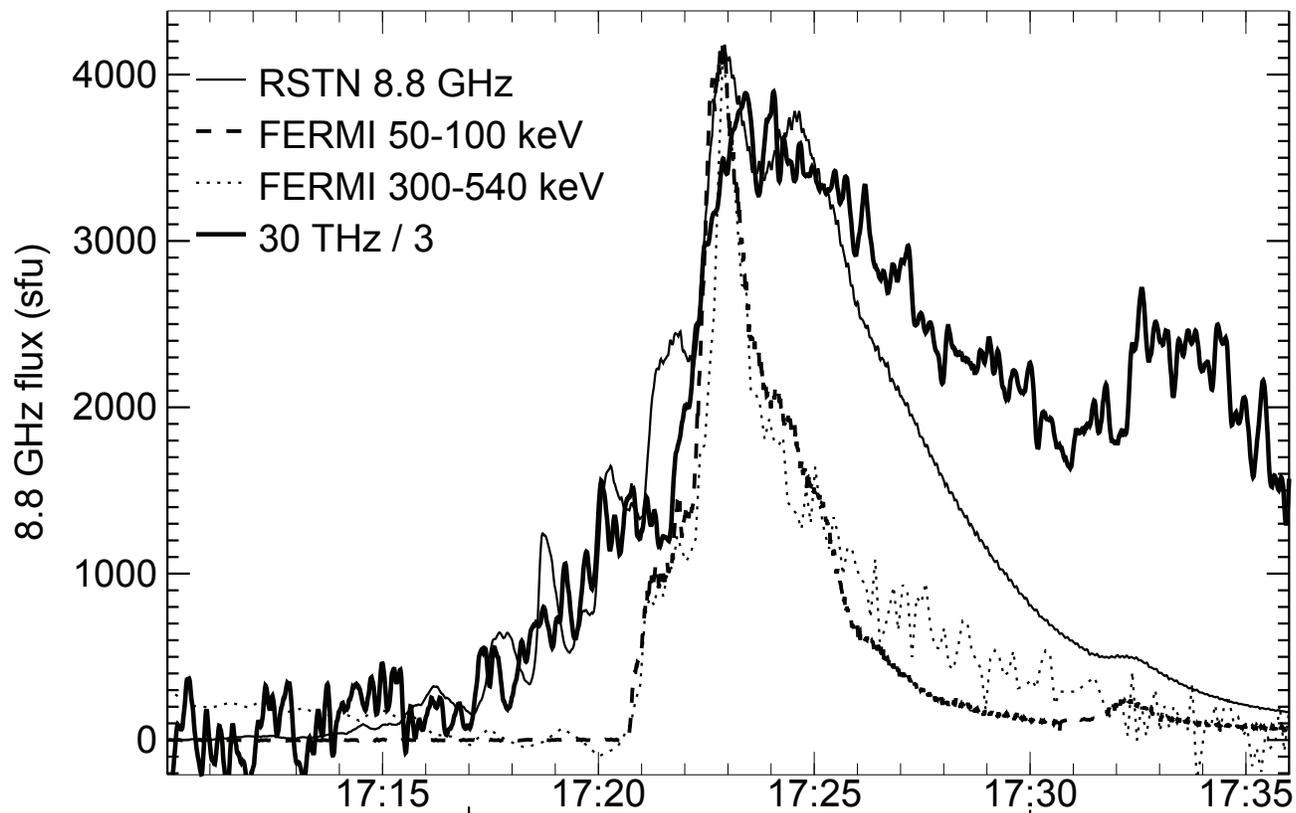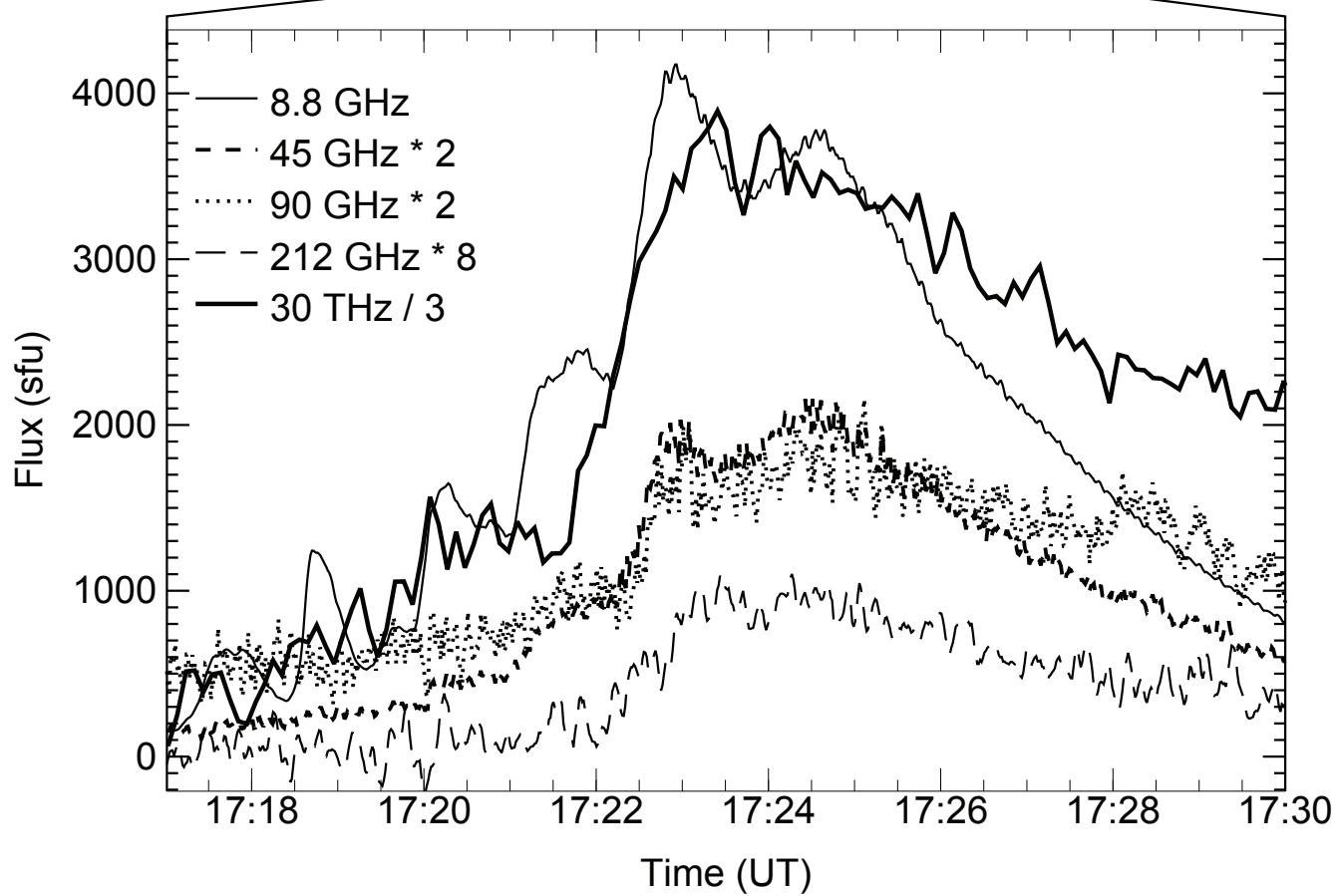

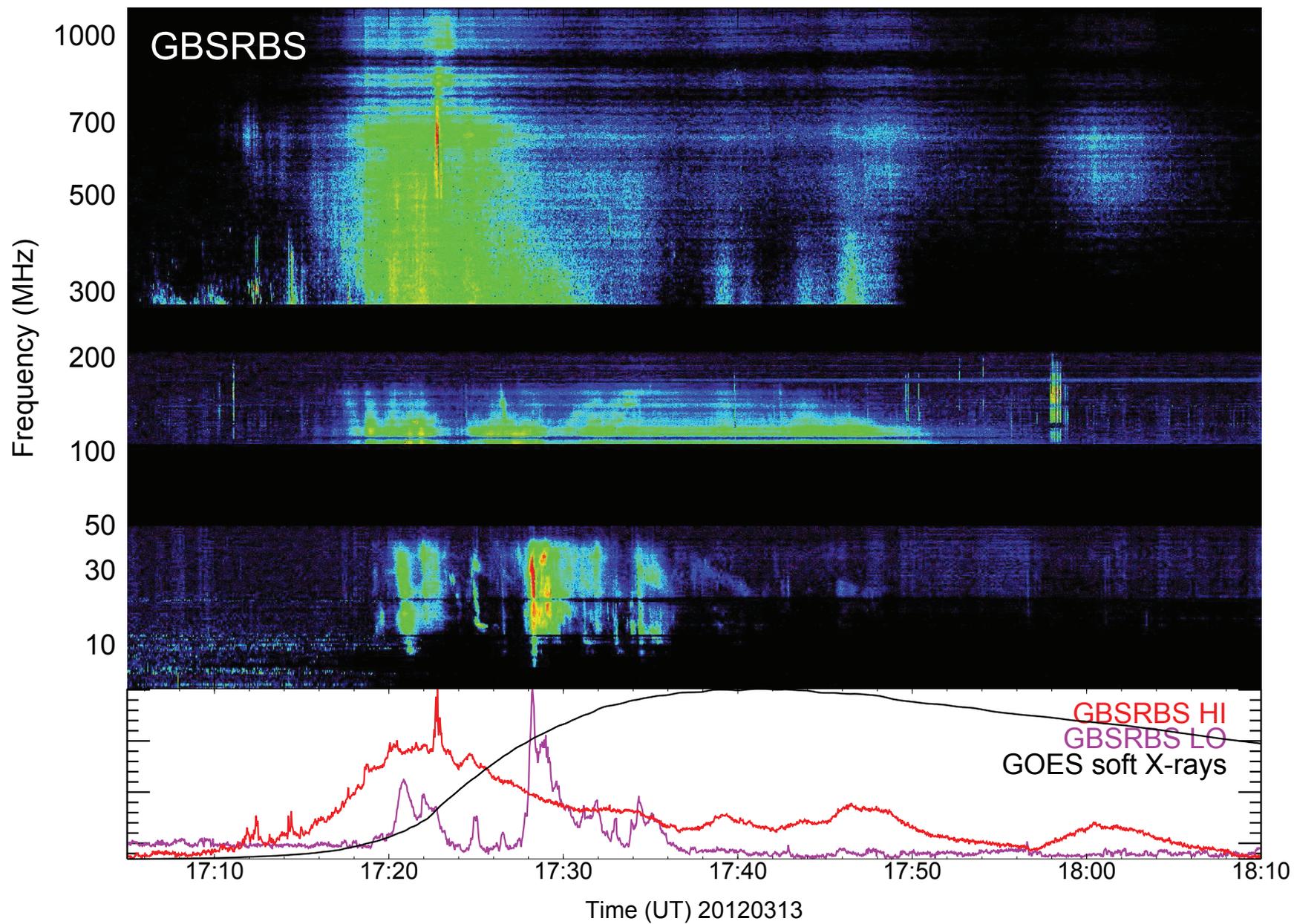

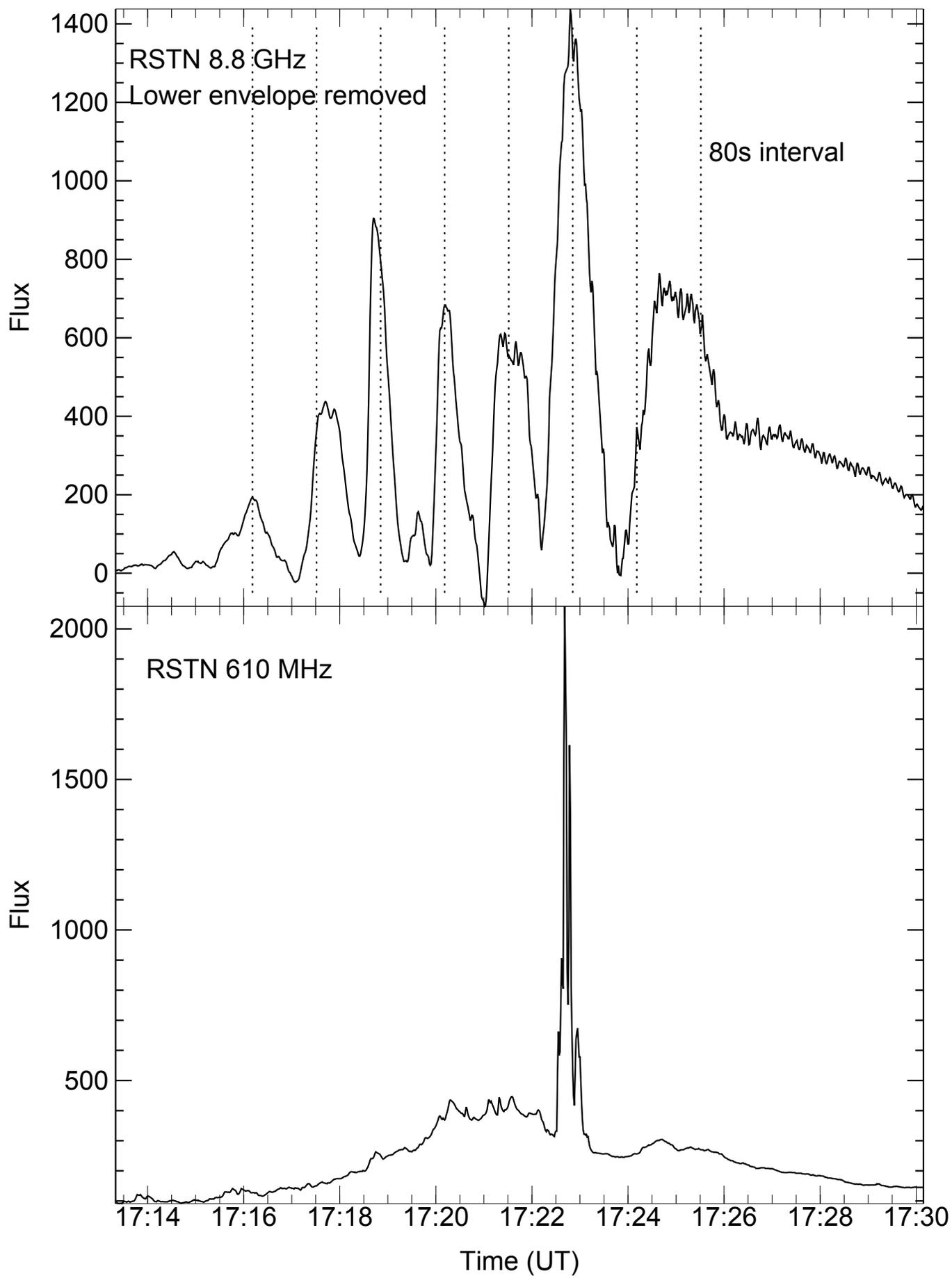

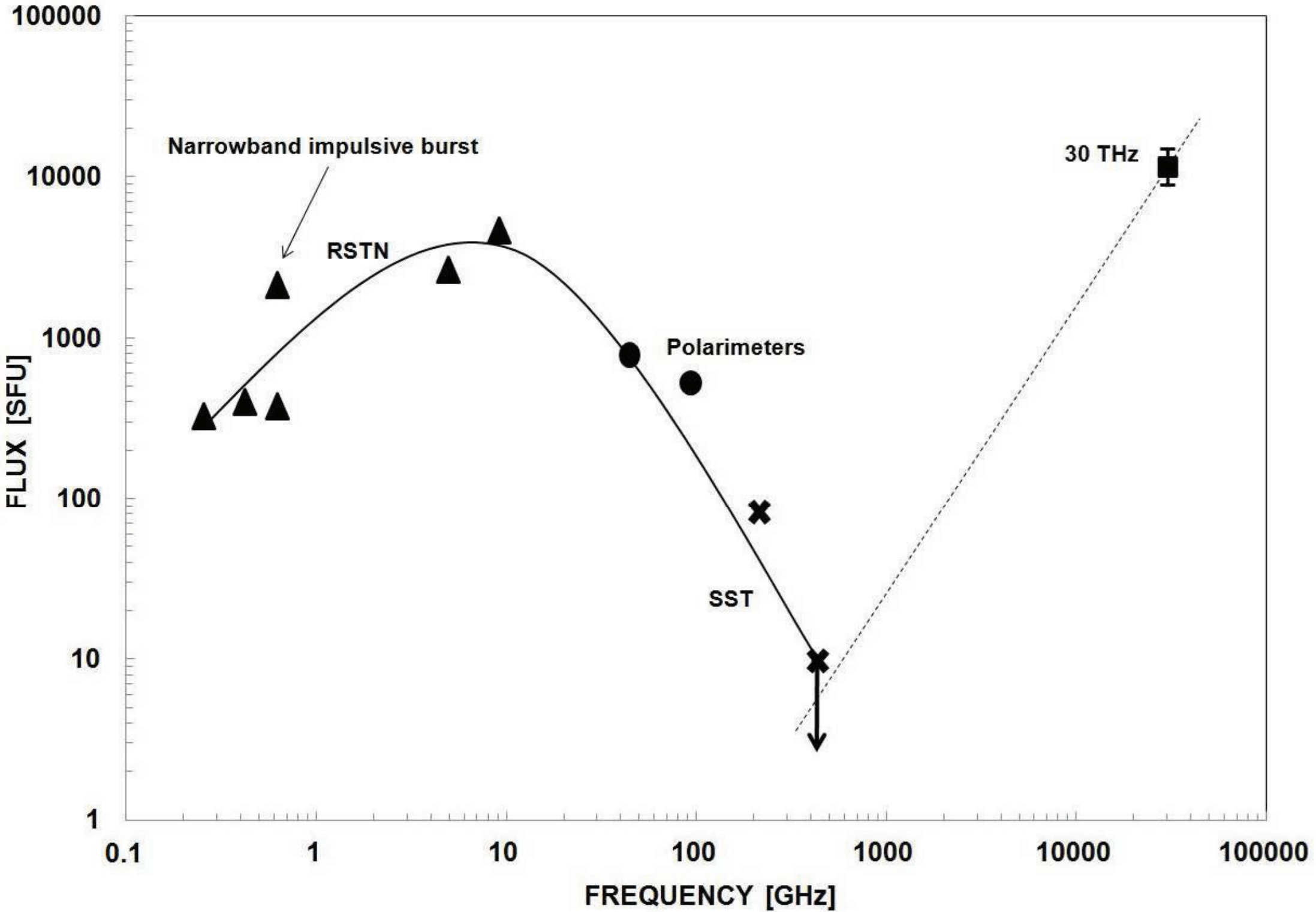

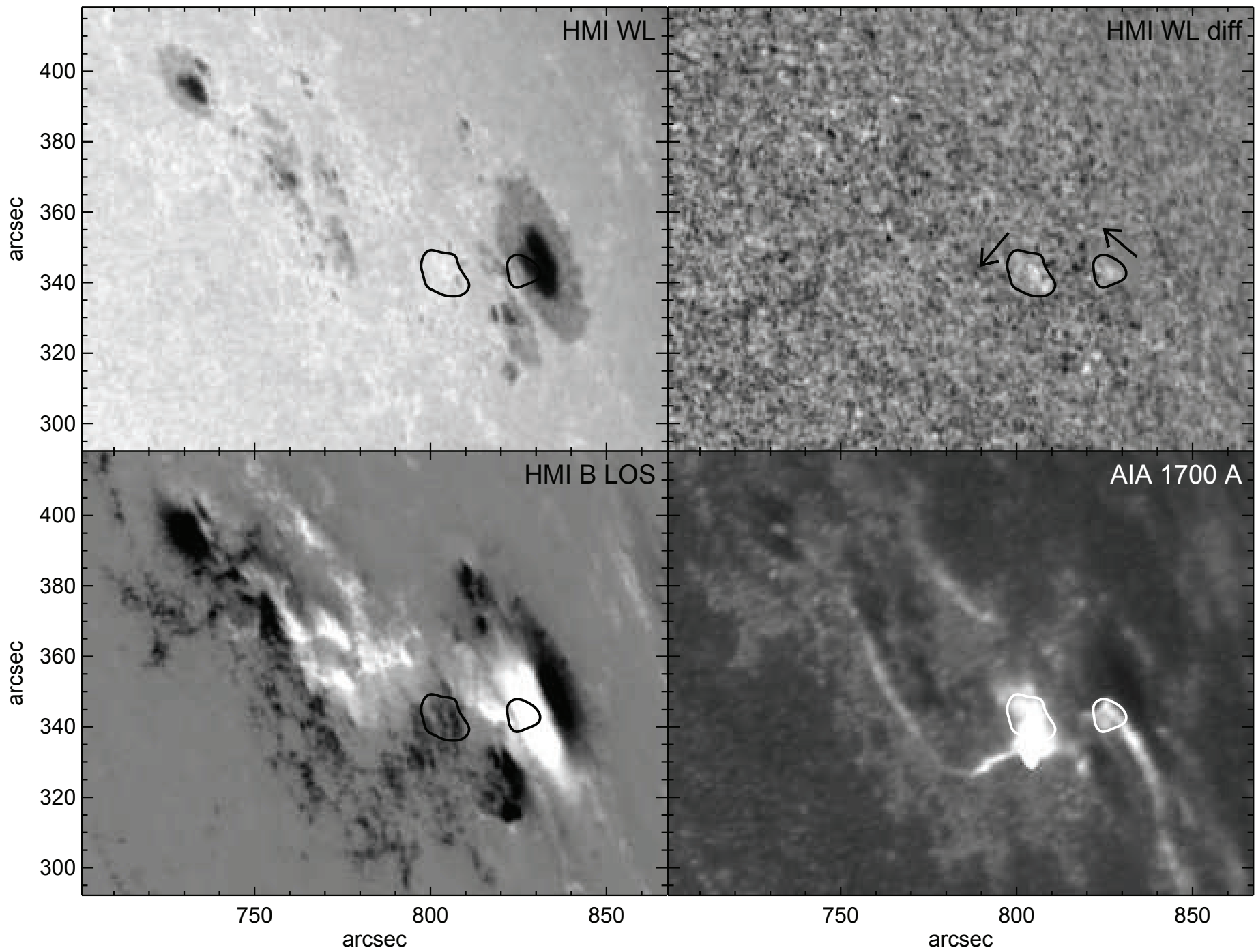

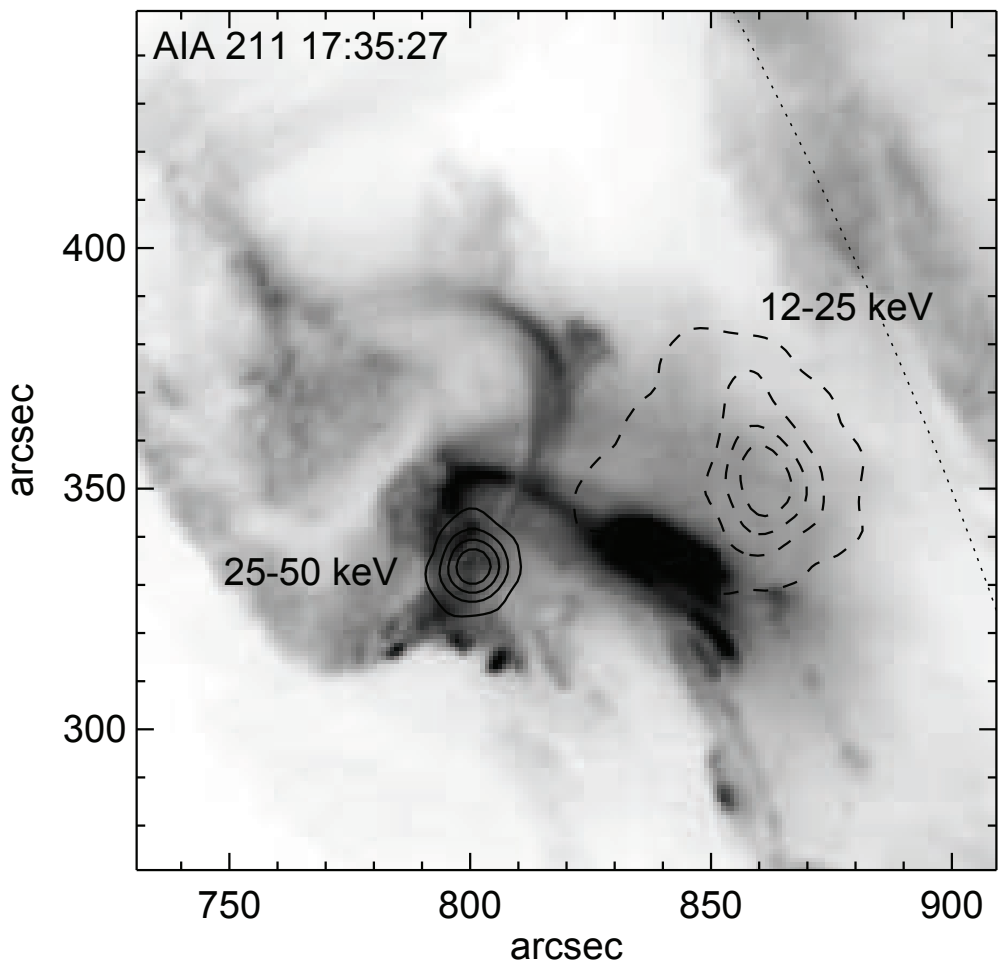